\documentclass[%
reprint,
showpacs,
 amsmath,amssymb,
 aps,
pra,
]{revtex4-1}
\usepackage{filecontents}
\usepackage{color,soul}
\usepackage[dvipsnames]{xcolor}
\usepackage{graphicx}
\usepackage{dcolumn}
\usepackage{bm}
\usepackage{hyperref}

\newcommand{\Er}{\ensuremath{^{166}{\rm Er}}}
\newcommand{\as}{\ensuremath{a_{\rm s}}}
\newcommand{\add}{\ensuremath{a_{\rm dd}}}
\newcommand{\nuax}{\ensuremath{\nu_{\rm axial}}}
\newcommand{\Bf}{\ensuremath{B_{\rm f}}}
\newcommand{\af}{\ensuremath{a_{\rm s}}}
\newcommand{\epsdd}{\varepsilon_{\rm dd}}
\newcommand{\br}{\ensuremath{{\boldsymbol r}}}
\newcommand{\bB}{\ensuremath{{\boldsymbol B}}}

\newcommand{\tho}{t_{\rm h}}
\newcommand{\tra}{t_{\rm r}}
\newcommand{\ttof}{t_{\rm ToF}}

\newcommand{\Nc}{N_{\rm core}}

\newcommand{\Nt}{N_{\rm th}}
\newcommand{\um}{\mu{\rm m}}
\newcommand{\dg}{\ensuremath{^{\rm o}}}
\newcommand{\parallelsum}{\mathbin{\!/\mkern-5mu/\!}}
\newcommand{\tsl}{t_{\rm slow}}
\newcommand{\tfa}{t_{\rm fast}}
\newcommand{\tD}{t_{\rm D}}
\newcommand{\td}{t_{\rm d}}
\newcommand{\ER}{E_{\rm R}}
\newcommand{\aSB}{a_{\rm SB}}
\begin{document}

\preprint{APS/123-QED}

\title{Quantum-fluctuation-driven crossover from a dilute 
Bose-Einstein condensate to a macro-droplet in a dipolar quantum 
fluid}
\author{L. Chomaz$^1$, S. Baier$^{1}$, D. Petter$^{1}$, M. J. Mark$^{1,2}$, F. W\"achtler$^3$, L. Santos$^3$, F. Ferlaino$^{1,2}$}
\email{Francesca.Ferlaino@uibk.ac.at}
\affiliation{%
 $^{1}$Institut f\"ur Experimentalphysik,Universit\"at Innsbruck, Technikerstra{\ss}e 25, 6020 Innsbruck, Austria\\
$^{2}$Institut f\"ur Quantenoptik und Quanteninformation,\"Osterreichische Akademie der Wissenschaften, 6020 Innsbruck, Austria\\
$^{3}$Institut f\"ur Theoretische Physik, Leibniz Universit\"at Hannover, Appelstr. 2, 30167 Hannover, Germany
}%

\date{\today}

\begin{abstract}

In a joint experimental and theoretical effort, we report on the formation of a macro-droplet state in an ultracold bosonic gas of erbium atoms with strong dipolar interactions. By precise tuning of the $s$-wave scattering length below the so-called dipolar length, we observe a smooth crossover of the ground state from a dilute Bose-Einstein condensate (BEC) to a dense macro-droplet state of more than $10^4$ atoms. Based on the study of collective excitations and loss features, we quantitative prove that quantum fluctuations stabilize the ultracold gas far beyond the instability threshold imposed by mean-field interactions.  
Finally, we perform expansion measurements, showing the evolution of the normal BEC towards a three-dimensional self-bound state and show that the interplay between quantum stabilization and three-body losses gives rise to a minimal expansion velocity at a finite scattering length.




\end{abstract}


\maketitle


\section{\label{Sec:Introduction}Introduction}
The extraordinary success of ultracold quantum gases largely relies on their simple description. Notably the actual inter-particle interactions, which might be complex or even unknown at short range, are very well captured in terms of simple mean-field (MF) potentials, proportional to the local particle density, $n$, and accounting for the average mutual effect of all neighboring particles\,\cite{Pitaevskii:2003}. 
In particular, the MF contact interaction is solely parametrized by the $s$-wave scattering length, $\as$, which in turn can be widely tuned 
 by means of Feshbach resonances (FRs)\,\cite{Chin:2010}.  
The MF treatment of a Bose gas leads to the famous Gross-Pitaevskii equation (GPE), which is very powerful in describing the underlying ground-state physics of a Bose-Einstein condensate (BEC), whereas the Bogoliubov-de Gennes (BdG) spectrum of collective modes thoughtfully accounts for excitations in the BEC \,\cite{Pitaevskii:2003}. 

Beyond the great achievements of dilute gases as testbed for MF theories, the quest for beyond-MF effects has triggered great interest in the ultracold community. The general question of how the many-body ground-state of bosons is modified by quantum fluctuations~(QFs) of elementary excitations was first addressed by Lee, Huang, and Yang~(LHY) in the 1950's~\cite{LHY:1957}. The so-called LHY term, which accounts for the first-order correction to the condensate energy, 
scales for a contact-interacting gas as $\as n \sqrt{n \as^3}$.
While in the weakly-interacting regime the effect of QFs is negligible and  difficult to isolate from MF contributions, it can be sufficiently amplified by increasing $\as$ via a FR. 
Based on this concept, recent beautiful experiments with alkali have observed clear shifts of the BdG spectrum and equation of state caused by the LHY term in strongly-interacting Fermi\,\cite{Altmeyer:2007,Shin:2008,Navon:2010} and Bose gases\,\cite{Papp:2008,Navon:2011}.

While in these measurements the role of QFs remains ancillary, it has been recently pointed out\,\cite{Petrov:2015} that in systems with various tunable interactions of different origin,  the overall MF energy can be made small and the LHY term becomes dominant. In this extreme regime, a novel phase of matter is expected to appear, namely a liquid-like droplet state. For purely contact-interacting gases, this situation is hardly realizable since it would require, for instance, Bose-Bose mixtures with coincidental overlapping FRs\,\cite{Petrov:2015}. In contrast, dipole-dipole interaction (DDI) offers genuinely this possibility in a single-component atomic gas by competing with the isotropic MF contact interaction\,\cite{Lahaye:2009,Baranov:2012}. 
 In the MF picture, a paradigm of the competition between DDI and contact interaction is embodied by the ability of quenching a dipolar BEC to collapse by varying $\epsdd =\add/\as$, where $\add=\mu_0\mu^2m/12\pi\hbar^2$ is a characteristic length set by the DDI, with $m$ the mass and $\mu$ the magnetic moment of the atoms\,\cite{Koch:2008,Lahaye:2008}. Here $\hbar$ stands for the reduced Planck constant and $\mu_0$ for the vacuum permeability. In general, because of the special geometrical tunability of DDI with the external trapping potential and dipole orientation, the stability and phase diagram remarkably depend on $\lambda=\nu_{\parallelsum}/\nu_{\perp}$, where $\nu_{\parallelsum}$ ($\nu_{\perp}$) is the trapping frequency along (perpendicular to) the dipole orientation\,\cite{Baranov:2012,Koch:2008,Metz:2009}.

In parallel, recent breakthrough experiments with an oblate dysprosium (Dy) dipolar BEC ($\lambda>1$) have shown that when quenching up $\epsdd$, the system, instead of collapsing, forms a metastable state of several small droplets\,\cite{Kadau, Igor}.
This observation has triggered an intense debate on the nature of such a state and its underlying stabilization mechanism\,\cite{Xi:2016,Bisset:2015,Blakie:2016,Waechtler:2016,Bisset:2016,Waechtler:2016b,Baillie:2016}. These works have eventually converged in confirming a scenario based on QFs\,\cite{Igor,Waechtler:2016,Bisset:2016} and meanwhile raise fundamental questions on the parameter space  of the effect, the single-to-multi droplet phase diagram, and the lifetime of the state. 
Aside from the Dy specificity,  will other dipolar systems, lying in different parameter ranges - \emph{e.g.} different dipolar strength ($\add$), geometry ($\lambda$) or atom number $N$ - form a droplet state? When does QF bear stabilization and when does this picture break down? The fundamental questions of the actual ground-state and of the resulting phase diagram as a function of $\lambda$, $\epsdd$ and $N$, remain widely unanswered beyond the quench dynamics and metastablity.

 The present work addresses these questions using a BEC of erbium (Er) atoms \footnote{Note that Er offers a system of intermediate dipolar strength ($\add=65.5a_0$) between Cr ($\add=15a_0$) and Dy ($\add=132a_0$).}, confined in a cigar-shaped harmonic trap with large elongation along the dipole orientation ($\lambda \ll 1$). In this condition, the DDI is strongly attractive, imposing a stringent condition to the system stability: already at $\epsdd\approx 1$, a dipolar BEC of about $10^5$ atoms should readily collapse within the MF description\,\cite{Lahaye:2009,Koch:2008}. 
 We explore the behavior of the dipolar gas by studying both the ground-state evolution and the dynamics when increasing $\epsdd$. Our work, based on complementary observations on density distributions, elementary excitations of the BdG spectrum, expansion dynamics, and lifetime of the dipolar quantum gas, show the existence of a crossover from a dilute BEC to a dense single droplet solution and quantitatively identify the driven role of QFs in the system dynamics.

\section{\label{Sec:ExpSetup}Experimental Procedures}
\begin{figure}
\includegraphics{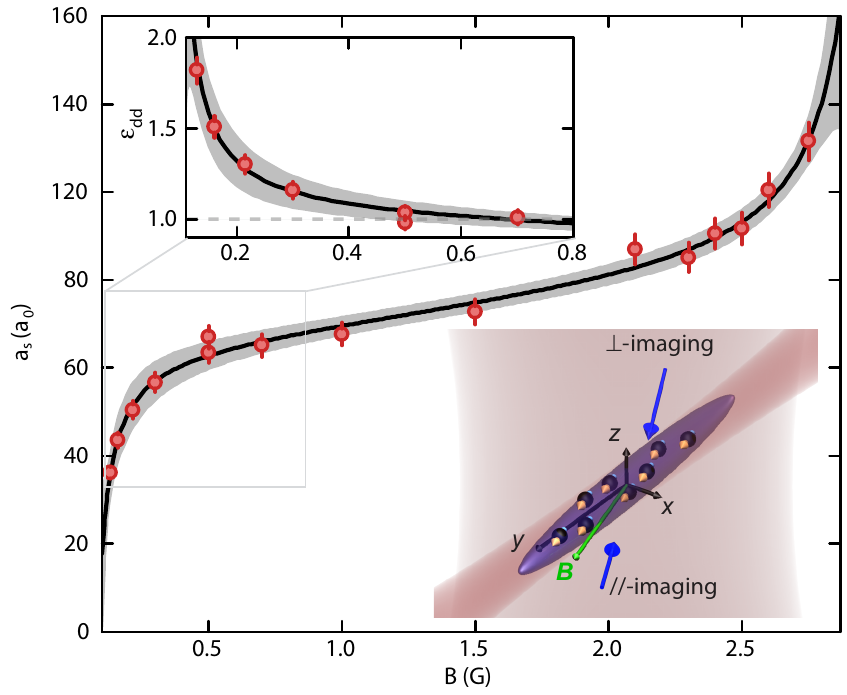}
\caption{\label{fig:setup}  (color online) {\bf Scattering length in $^{166}$Er}:
$\as$
as a function of $B$. The data points (circles) are extracted from spectroscopic measurements in a lattice-confined gas and the solid line is a fit to the data with its statistical uncertainty (grey shaded region\,\cite{suppmat}). 
(Upper inset) Zoom in of  $\varepsilon_{\rm dd}$ as a function B.  The grey dashed line marks $\varepsilon_{\rm dd}=1$; see also the other figures.
The lower inset illustrates the geometry of our experimental setup, the relevant axes ($x$,$y$,$z$), the crossed-optical-dipole-trap beams (shaded regions), the magnetic field orientation (green arrow), and the $\parallelsum$- and $\perp$-imaging view axes (blue arrows).
}
\end{figure}

The atomic properties of Er offer a  privileged platform to explore a variety of interaction scenarios. Beside its strongly magnetic character and its many FRs\,\cite{Frisch:2014}, Er has several stable isotopes. This feature adds an important flexibility in term of the choice of the background $\as$\,\cite{us:inPrepapration}. 
In our early work on Er BECs, we employed the $^{168}{\rm Er}$ isotope, which has a background $\as$ of about twice as large as the dipolar length, $\add=65\,a_0$\,\cite{Aikawa:2012,Baier:2015}. 

In the work reported here, we produce 
and use a BEC of 
\Er \ in the lowest internal state. 
This isotope provides us with two major advantages. First, its  background $\as$ is comparable to its dipolar length, $\add= 65.5\,a_0$, realizing $\epsdd=\add/\as\approx 1$ without the need of Feshbach tuning. Second, \Er \ features a very convenient FR at ultralow magnetic-field values, $B$. 
To precisely map $\as$ as a function of $B$, we use a spectroscopic technique based on the measurement of the energy gap of the Mott insulator state in a deep three-dimensional optical lattice\,\cite{Mark:2011, Baier:2015}. A detailed description is given in the Supplementary Material\,\cite{suppmat}.
Between $0$ and $3\,$G we observe a smooth variation of $\as$, which results from two low lying FRs whose centers are fitted to $0.05(5)$\,G and $3.0(1)$\,G respectively; see Fig.\,\ref{fig:setup}. 
This feature gives an easy access into the $\epsdd>1$ regime, allowing variation of $\epsdd$ from $0.70(2)$ to $1.58(18)$ by changing $B$ from 2.5\,G to 0.15\,G; see Fig.\,\ref{fig:setup} upper inset.   
By fitting our data\,\cite{Chin:2010}, we extract 
$\as(B)$ valid for $B$ in the $[0.15,2.5]-$G range, which we will use along this manuscript\,\cite{suppmat}.

We achieve Bose-Einstein condensation of \Er \ using an all-optical scheme very similar to Ref.\,\cite{Aikawa:2012} with  cooling parameters optimized for \Er\,\cite{suppmat}. In brief, we drive forced evaporative cooling at a magnetic field $B=1.9\,{\rm G}$, corresponding to $\as=81(2)a_0$ ($\epsdd=0.81(2)$). In this phase $B$ is oriented along the vertical $z$ axis. At the end of the evaporation, we obtain a BEC of $N=1.2 \times 10^5$ atoms with a condensed fraction above $80\%$.  

To reach the $\lambda\ll 1$ regime, we slowly modify, in the last step of the evaporation, the confining potential 
to the final cigar shape, with typical frequencies $(\nu_x,\nu_y,\nu_z)=(156(1),17.2(4),198(2))\,$Hz. Simultaneously, we decrease $B$ to 0.8\,G ($\as=67(2)\,a_0$) and then change the magnetic-field orientation to the weak trapping axis ($y$) while keeping its amplitude constant\,\cite{suppmat}.  
Finally, we ramp  $B$ to the desired target value  (and equivalently $\as$)  in $\tra$\,\cite{suppmat}, hold for a time $\tho$ and perform absorption imaging of our gas after a time-of-flight (ToF) of $t_{\rm ToF}$.
 Two imaging setups are used in order to measure either the density distribution integrated along the dipoles ($\parallelsum$-imaging) or perpendicular to them ($\perp$-imaging)\,\cite{suppmat}.
Figure\,\ref{fig:setup} (lower inset) illustrates the final geometry of our system with $\nu_{\parallelsum}=\nu_y$, $\nu_{\perp}=\sqrt{(\nu_x^2+\nu_z^2)/2}$ giving $\lambda=0.097(3)$, and defines the relevant axes.




\begin{figure}
\includegraphics{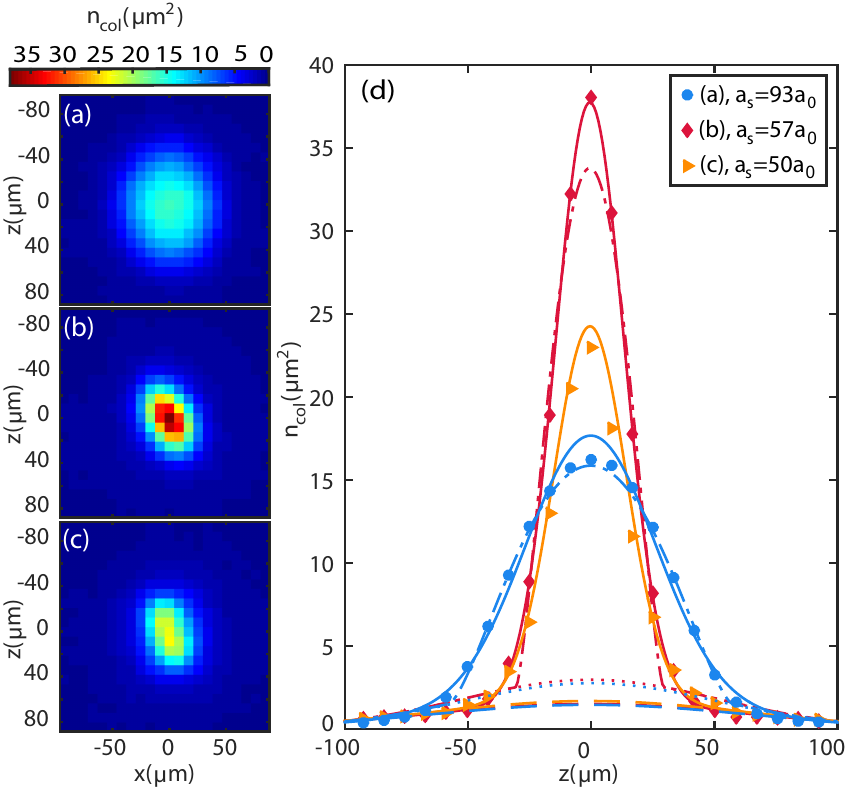}
\caption{\label{fig:density_profile}  
(color online) {\bf Density profiles in the BEC-to-droplet crossover }: (a-c) 
2D column density distributions probed  with $\parallelsum$-imaging and (d) corresponding central cuts along the $x=0$ line (dots) after a ramp of $\tra=10\,$ms and holding of $\tho=6\,$ms to different $\af$: $93\,a_0$ ((a) and circles in (d)), $57\,a_0$  ((b) and diamonds in (d)),  $50\,a_0$ ((c) and triangles in (d)). Each  distribution is obtained by averaging 4 absorption images taken after $\ttof=27\,$ms. 
In (d), the lines show the central cuts of the 2D bimodal fit results. The solid (resp. dashed-dotted) lines show the fit to a two-Gaussian (resp.  MF TF plus Gaussian) distribution and the dashed (resp. dotted) lines show the broad thermal Gaussian of the corresponding fits. 
For visibility, not all MF TF plus Gaussian bimodal fits are shown.}
\end{figure}

Here, we explore the 
properties of the system when the DDI is made attractive enough to overcome the repulsive MF contact interaction ($\lambda\ll 1$, $\epsdd > 1$), after adiabatically changing ($\tra\geq 45\,$ms) or quenching ($\tra= 10\,$ms) $\as$ to 
its target value\,\cite{suppmat}. 
For $\tra \geq 45\,$ms, the system evolves following its ground state and gives access to the slow dynamics, whereas for the $\tra= 10\,$ms case we can probe the fast dynamics and study the relaxation towards an equilibrium.
The key question is whether QFs protect the system from collapsing.
Indeed, in this regime, the MF treatment would imply that the attractive BEC becomes unstable, leading to a two-fold dramatic consequence\,\cite{Pitaevskii:2003}. First, some modes of the BdG spectrum acquire complex frequencies. Second, in a trap, the density distribution of the cloud undergoes a marked change on short time scales ($\leq 1/\nu_\perp$), described as a ``collapse", which can develop into a rapid loss of coherence\,\cite{Roberts:2001,Koch:2008}, 
and pattern formations, such as anisotropic atom bursts ("Bosenova") and special $d$-wave-type structures, as observed in rubidium \,\cite{Donley:2001} and dipolar gases of chromium\,\cite{Koch:2008,Lahaye:2008}, respectively. This fast dynamics has been proved to be  well encompassed by GPE simulation \cite{Ueda:2003,Lahaye:2008,Metz:2009}. 


\section{\label{Sec:DensityDistr}Density Distribution}

In a first set of experiments, we study the evolution of the ToF density distribution of our dipolar Er BEC with $\af$. Figure \ref{fig:density_profile}\,(a-c) shows the 2D column density profiles acquired with $\parallelsum$-imaging for $\tra=10\,$ms and $\tho=6\,$ms ($\gtrsim 1/\nu_\perp$), and for different $\af$. To analyze our data, we fit the 2D density profiles to bimodal distributions. We illustrate their results on the central cuts  ($x=0$) in Fig.\,\ref{fig:density_profile}(d). For $\af>\add$ (a), the density distribution follows the one of a dilute BEC with the expected MF Thomas-Fermi (TF) profile on top of a broad Gaussian distribution, accounting for the thermal atoms. 
When lowering $\af$ below $\add$, we observe a drastic change of the density profile marked by the shaping of a narrower and denser central core (b-c), whose profile starts to deviate from the usual MF TF one. 
Indeed, the $\parallelsum$-images are better described by 
the sum of two Gaussian functions already for $\af<70\,a_0$ as the fit to this distribution 
gives a smaller residue than the MF TF plus Gaussian bimodal fit; see Fig.\,\ref{fig:density_profile}\,(d). 
In contrast with the evolution of the central peak, the distribution of the thermal atoms, encompassed by the broad Gaussian function, remains  mainly unaffected by the change of $\as$. This remarkable behavior evidences an absence of a sizable heating and an apparent decoupling of the evolution of the coherent and thermal parts.   
The central peak remains visible down to $\af \sim 50\,a_0$ ($\epsdd \sim 1.3$) (c), and exhibits a  long-lived character from several tens to hundreds of ms. 
As discussed in details later and suggested in Ref.\,\cite{Baillie:2016}, three-body (3B) collisions regulate the lifetime of the high-density component, as this loss mechanism exhibits a high-power dependence on the density.
We note that we observe a similar qualitative behavior of the density distribution when using an adiabatic ramp of $\af$. 
However, in this case 3B losses already sets in during the ramp reducing the importance of the central peak.

Being a smoking-gun for long-range phase coherence, the survival of a bimodal profile in the ToF distribution
suggests a persistent coherent behavior far beyond the MF instability threshold. 
The absence of a collapse
advocates the outbreak of an additional stabilization mechanism. 

\section{\label{Sec:CollectiveOsc}Collective Oscillation}

\begin{figure}
\includegraphics{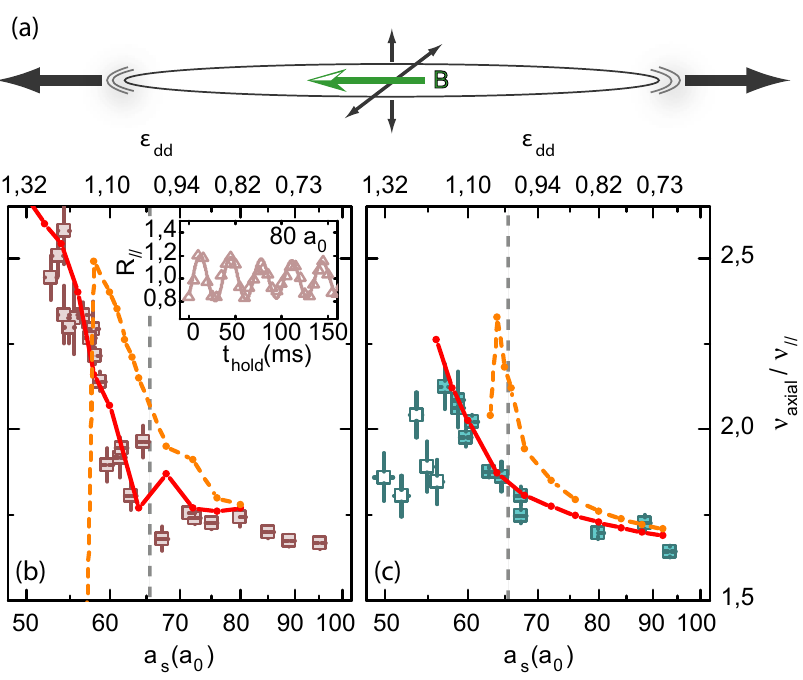}
\caption{\label{fig:modes} 
(color online) {\bf Axial mode}:  (a) Illustration of the axial mode in our experimental setup. The black arrows sketch the oscillations of the widths of the coherent gas along the characteristic axes of the trap, with weights indicating their relative amplitudes. (b-c) Measured $\nuax/\nu_{\parallelsum}$ (squares) as a function of $\af$   together with the theoretical predictions, including (solid line) or not (dashed lines) the LHY term for $\tra=100\,$ms (b) and $\tra=10\,$ms (c). In the latter case, quenches to small $\as$ excite  higher-energy modes and the system is not anymore in the linear response regime  (open squares). The inset in (b) exemplifies a measurement of $R_{\parallelsum}$ (triangles) and its fit to a damped-sine (solid line) for $\af=80\,a_0$. We typically fit 4 to 5 oscillations for all our $\as$.
}
\end{figure}


In a second set of experiments, we unveil the origin of stabilization mechanism by studying the elementary excitations of the coherent cloud. This is a very powerful probe of the fundamental properties in quantum degenerate gases~\cite{Pitaevskii:2003,Grimm:2007}. In particular, collapse is intimately related to the softening of some collective modes at the MF-instability threshold.
%
We focus here on the axial mode, which is the lowest-lying excitation in the system.
It corresponds to a collective oscillation of the condensate length  along $y$~($R_{\parallelsum}$) with frequency $\nu_{axial}$. The axial oscillation comes along with a smaller-amplitude oscillation of the radial sizes in phase-opposition; see Fig.\,\ref{fig:modes}(a).  As a result, this mode has a mixed character between a compression and a surface mode~\cite{Pitaevskii:2003}. The compression character is particularly relevant since it involves a change in the density and it is therefore sensitive to the LHY corrections \cite{PhysRevLett.81.4541}.

We excite the axial mode either by ramping $B$ during the final preparation stage or by transiently increasing the power of the vertical optical dipole trap (ODT) beam, after ramping $B$ to $\Bf$. Here  $\nu_{\parallelsum}$ is abruptly changed from 17\,Hz to typically 21\,Hz, kept at this higher value for 8\,ms and finally set back to 17\,Hz. 
Following the excitation, we let the cloud evolve for a variable $\tho$ and 
 image its ToF density distribution with $\perp$-imaging. 
 To extract $\nuax$, we probe the axial width $R_{\parallelsum}$ of the central coherent component of the gas\,\cite{suppmat} with $\tho$ and fit it to a damped-sine; see inset in Fig.\,\ref{fig:modes}(b). 

Figure \ref{fig:modes} shows the observed $\nuax$ normalized to the  trapping frequency $\nu_{\parallelsum}$ \footnote{$\nu_{\parallelsum}$ is determined from the center-of-mass oscillations along $y$, which are simultaneously excited with the axial breathing mode.} as a function of $\af$ for  adiabatic 
(b) and non adiabatic 
(c) ramps. Both cases exhibit a similar qualitative behavior. For $\af>\add$, the oscillations shows a smooth dependence on $\epsdd$ with $\nuax$ increasing by about $5\%$ around $1.70\,\nu_{\parallelsum}$. When lowering $\as$, the oscillation of the coherent part remains visible well below the $\epsdd= 1$ threshold and $\nuax$ exhibits a marked increase. $\nuax/\nu_{\parallelsum}$ grows up to $2.6(1)$ at $\af=54\,a_0$ for $\tra=100\,$ms (b). 
For $\tra=10\,$ms (c), $\nuax/\nu_{\parallelsum}$ first increases similarly to the adiabatic case (b), reaches a maximum of $\sim 2.13(7)$ at $57\,a_0$ ($\epsdd=1.15$), and finally decreases for even smaller $\af$ (open squares). The latter behavior can be explained by the fact that the larger quenches in the interaction excites additional high-energy modes while it drives the system away from the linear response regime \footnote{We indeed observe high-energy mode in which the high-density core splits and recombine in a damped oscillating manner. }. 

\section{\label{Sec:Theory}Theory}

To account for our observation and discern between the MF instability picture and QF mechanisms, we develop a beyond-MF treatment of our system at $T=0$.  The coherent gas is described here by means of the generalized non-local nonlinear-Schr\"odinger equation~(gNLNLSE), which includes the first-order correction from QF, i.\,e.\,the LHY term, and 3B  loss processes. The gNLNLSE reads as ~\cite{Waechtler:2016,Baillie:2016}
\begin{equation}
\label{eq:H}
{\mathrm i}\hbar\frac{\partial\psi}{\partial t}=\left[\hat{H}_0 +\mu_\text{MF}(n,\epsilon_{dd})+\Delta\mu(n,\epsilon_{dd})-{\mathrm i}\hbar \frac{L_3}{2}n^2 \right]\psi,
\end{equation}
where $\hat H_0=\frac{-\hbar^2\Delta}{2m}+V(\br)$ is the non-interacting Hamiltonian and $V(\br)=2\pi^2m \sum_{\eta=x,y,z}\nu_\eta^2 \eta^2$ the harmonic confinement with $\eta=x,y,z$.
The MF chemical potential, $\mu_\text{MF}(n({\mathbf r}),\epsilon_{dd})=gn({\mathbf r})+\int d^3 r' V_{\rm dd}({\mathbf r}-{\mathbf r}')n({\mathbf r}')$, results from the competition between
short-range interactions, controlled by the coupling constant $g=4\pi \hbar^2 \as/m$, 
and the DDI term with $V_{\rm dd}({\mathbf r})=\frac{\mu_0\mu^2}{4\pi r^3}(1-3\cos^2\theta)$ and
$\theta$ the angle sustained by ${\mathbf r}$ and the dipole moment ${\mathbf \mu}$. Here $n({\mathbf r})=|\psi({\mathbf r})|^2$. 
The beyond-MF physics is encoded in the LHY term, leading to an additional repulsive term in the chemical potential , $\Delta \mu (n,\epsilon_{dd})=\frac{32}{3\sqrt{\pi}}gn\sqrt{na^3}F(\epsilon_{dd})$. The function
$F(\epsilon_{dd})=\frac{1}{2}\int d\theta_k \sin{\theta_k} (1+\epsilon_{dd}(3\cos^2\theta_k-1))^{5/2}$ 
is obtained  from the LHY correction in homogeneous 3D dipolar BECs~\cite{Pelster:2011,Pelster:2012} using local-density approximation \footnote{Our experiments, performed in cigar-shape traps with relatively large
condensates, are well within the validity regime of this approach, which has been discussed in detail in Refs.\,\cite{Waechtler:2016,Waechtler:2016b}.}. The last non-Hermitian term in Eq.\,\eqref{eq:H} accounts for 3B loss processes\,\cite{Kagan:1998}. In our calculations, we use the experimentally determined values of the 3B recombination rate of the condensate, $L_3 (\as)$ \cite{suppmat}.

As discussed in Refs.\,\cite{Waechtler:2016b,Baillie:2016}, due to the repulsive LHY term, Eq.\,\eqref{eq:H} sustains stable ground-state solutions for any $\as$ and $\lambda$. 
For pancake traps ($\lambda>1$), the solution of Eq.\,(\eqref{eq:H}) is not unique. The phase diagram reveals three types of solutions: the one of a dilute BEC, a single droplet solution, and a third one, which separates the previous two phases, that corresponds to a metastable region of multi-droplet states. The latter has been observed in Dy experiments \cite{Kadau}. However, the single droplet solution appears difficult to access because of the overhead multi-droplet state and the stringent 3B loss mechanisms.
Remarkably, in cigar-shaped traps ($\lambda<1$) Eq.\,\eqref{eq:H} has only one possible solution. In the $\epsdd$ parameter space, the corresponding wave function exhibits a smooth crossover from a dilute BEC to a single, high-density, macro droplet solution for increasing $\epsdd$. It is worth to notice that the crossover physics, e.\,g.\,the formation and lifetime of the droplet state, is expected to crucially depend on the 3B collisional processes. In the following, we will concentrate on the $\lambda<1$ case, which corresponds to our experimental setting.

The continuous and smooth change of the static properties of the system with increasing $\epsdd$ is consistent with our observations on the evolution from a dilute into an high-density state; see Fig.\,\ref{fig:density_profile}.   

Based on Eq.\,\eqref{eq:H}, we theoretically investigate the dynamics of the coherent gas. In order to compare as close as possible to our experimental results, we precisely account for the experimental sequence by performing real time evolution (RTE) starting from the ground state of Eq.\,\eqref{eq:H} at $\as=67\,a_0$ with $N=1.2\times 10^5$ atoms.  We simulate a linear ramp in $\as$ from $67\,a_0$ to a variable final value of $\as$ in $\tra$, followed by a compression of the axial trap  from $\nu_{\parallelsum}=17.3$ Hz to  $21$ Hz during $8$ ms. We record then the axial width from the standard deviation of $n(\br)$, $\sigma_y=\sqrt{\langle y^2 \rangle}$, as a function of the subsequent holding time. The evolution of $\sigma_y$ is well fit by a sinusoidal function, whose frequency constitutes our theoretical prediction of $\nuax$.

In Fig.\,\ref{fig:modes}, we present our calculations with and without the LHY term. From a direct comparison with the experiments, 
one observes an excellent agreement of the theory including QFs, thus ruling out the MF scenario. In quench experiments (c), the effect of QFs is particularly evident since 3B losses do not have enough time to fully develop at the end of $\tra$, leaving the system in the high-density regime. In this condition, QFs play the crucial role of stabilizing the system, and drive the formation of a special coherent state, namely a single macro-droplet\,\cite{Waechtler:2016,Bisset:2016,Waechtler:2016b,Baillie:2016}.
This provides a qualitatively different phase diagram than the one from the MF treatment, which  predicts a collapse at $\as\approx 64\,a_0$.
For $\tra=100$\,ms, a more stringent interplay between QFs and 3B losses arises, both mechanisms being able to drive the system out of the instability region. Although the system is expected to be stable even within the MF description down to $\as\approx 57 a_0$, the effect of QFs is visible in a sizable shift of $\nuax$, which better matches the experimental data; see Fig.\,\ref{fig:modes}(b).

\section{\label{Sec:Losses}Lifetime of the high-density state}

\begin{figure}
\includegraphics{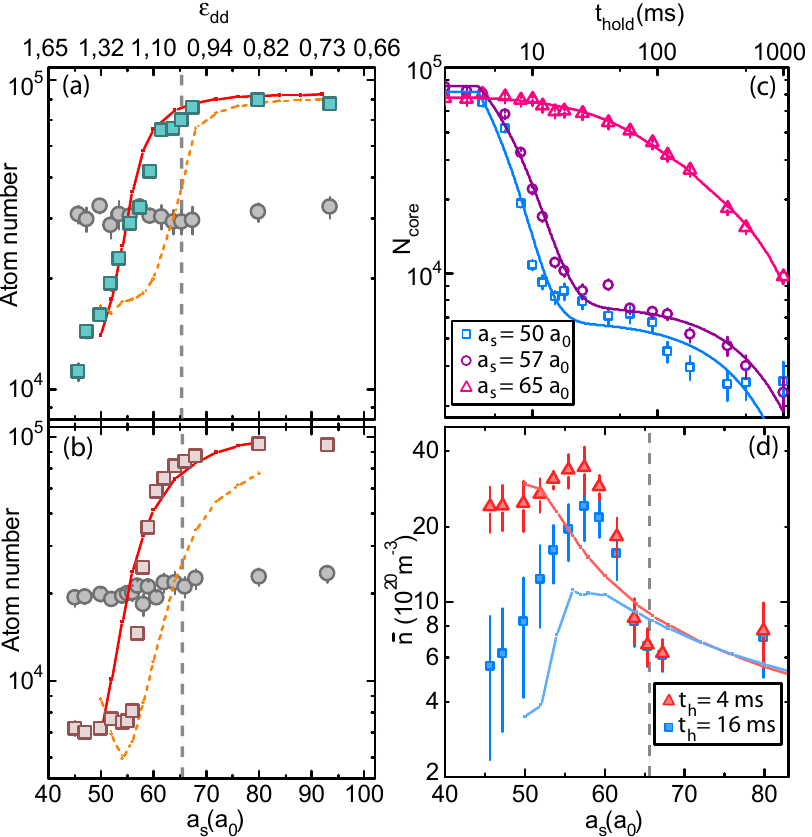}
\caption{\label{fig:losses} (color online) {\bf Lifetime and in-situ density of the high-density core}. 
(a-b) Measured $\Nc$ (squares) and $\Nt$ (circles) versus $\as$ after (a) a non-adiabatic ($\tra=10\,$ms, $\tho=8\,$ms) and (b) an adiabatic  ($\tra=45\,$ms, $\tho=0\,$ms) ramps. The data show a better agreement with the theory with the LHY term (solid line) as compared to the MF theory (dashed line). 
(c)  Time decay of $\Nc$ for $\af=65 a_0$ (triangles), $57 a_0$ (circles), and $50 a_0$ (squares) after quenching $\as$ ($\tra=10\,$ms). We fit a double exponential function to the data (solid lines) \cite{suppmat}. (d) From the fit, we deduct the mean in-situ density of the core, $\bar{n_c}$ (see text), for  $\tho=4\,$ms (triangles) and $16\,$ms (squares) and as a function of  $\af$. The error bars include the statistical errors on the fit and on $L_3$. The solid lines show results of the RTE including the LHY correction for the $\tho=0\,$ms (red),  $\tho=25\,$ms (blue).
}
\end{figure}



To further investigate the respective roles of 3B losses and QFs, we study the time evolution of the atom number of both the central core ($\Nc$) and thermal ($\Nt$) components along the BEC-to-droplet crossover.
Since in the droplet regime the core density $n_{\rm core}(r)$ dramatically increases, 3B losses are expected to play an important role even for moderate and low values of $L_3$ \cite{Baillie:2016}.
Notwithstanding, 3B losses and QFs exhibit different power dependencies on $n(r)$ (see Eq.\,\eqref{eq:H}) and thus the atom-loss dynamics should disclose the competition between diluteness and a droplet character.

Figure \ref{fig:losses}\,(a-b) shows $\Nc$ and $\Nt$, extracted from the double Gaussian fit  as a function of $\as$ after a non-adiabatic (a) and adiabatic (b)  change of $\as$. Both cases show a similar evolution. When lowering $\as$, $\Nc$ is first constant for $\as > \add$, then shows a sharp drop starting around $\as \sim \add$ and finally  saturates for lower $\as$. We note that in the non-adiabatic case, $\Nc$ decreases slower as compared to the adiabatic case because of the shorter timing involved.
Remarkably, $\Nt$ remains mainly unaltered over the whole range of $\as$ and the whole system does not show any appreciable heating.  This suggests that the condensed atoms, which are ejected from in the core, leave the trap instead of being transfer to the thermal component, confirming a picture in which the thermal and the condensed component have uncoupled dynamics.


We now compare the experiment with the theory, which, as previously, precisely accounts for the experimental sequence and its timing by performing RTE along Eq.\,\eqref{eq:H}.  We compute here the final atom number $N=\int n({\br}) {\rm d}^3 r$ as a function of $\as$ with and without the LHY term. 
Remarkably, the observed evolution of $\Nc$ is very well reproduced by our beyond-MF calculations (solid lines), whereas in absence of the LHY stabilization, the calculations predict losses in the condensed core to occur at values of $\as$ too large  compared to the measured ones; see Fig.\,\ref{fig:losses}(a,b,d).

Finally, we investigate the in-trap time evolution of $\Nc$ after quenching $\as$ in the droplet regime; see Fig.\,\ref{fig:losses}\,(c). Our measurements reveal three different timescales for the losses. At very short $\tho$ ($\approx 0-8$\,ms), $\Nc$ is constant and the droplet state preserves its high density. It follows a fast decay dominated by 3B ($\approx 8-25$\,ms), having the effect of ejecting atoms from the core until $\Nc$ decreases slightly below $10^4$ atoms. At this point, the loss dynamics substantially slows down ($\approx 25-1000$\,ms).
Using the relation $\frac{1}{\Nc}\frac{{\rm d}\Nc}{d\tho}=- L_3 \bar{n}^2$, we extract the mean in-situ density $\bar{n}=\sqrt{\langle n(r)^2 \rangle}$ of the high-density component in the BEC-to-droplet crossover. Figure\,\ref{fig:losses}\,(d) shows our results for $\tho=4,\, 16\,$ms. We observe a prominent increase of $\bar{n}$ across the threshold $\af \sim \add$, and a surviving 
high density state deep into the  MF instability regime. 
The formation of the droplet state is particularly visible for the $\tho=4$-ms case. Here, $\bar{n}$ grows from $6.2(9)\times 10^{20}\,{\rm m}^{-3}$ at $\af=67\,a_0$ to a maximum of $35(7)\times 10^{20}\,{\rm m}^{-3}$ at $\af=57\,a_0$, while it is slightly reduced to $\sim 24\times 10^{20}\,{\rm m}^{-3}$ at $\af\sim 46\,a_0$. These observations are qualitatively reproduced by our simulations including the LHY correction and 3B losses.

Our results together with the good agreement between theory and experiments provide an alternative confirmation of the central role of beyond-mean-field physics. The lifetime of the high-density core reveals, on the one hand, the activation of the LHY term and the crossover toward a dense droplet state, and on the other hand the counteracting role of 3B losses in regulating the maximum density in the droplet regime.


\section{\label{Sec:ExpansionDynamics}Expansion Dynamics}
\begin{figure}
\includegraphics{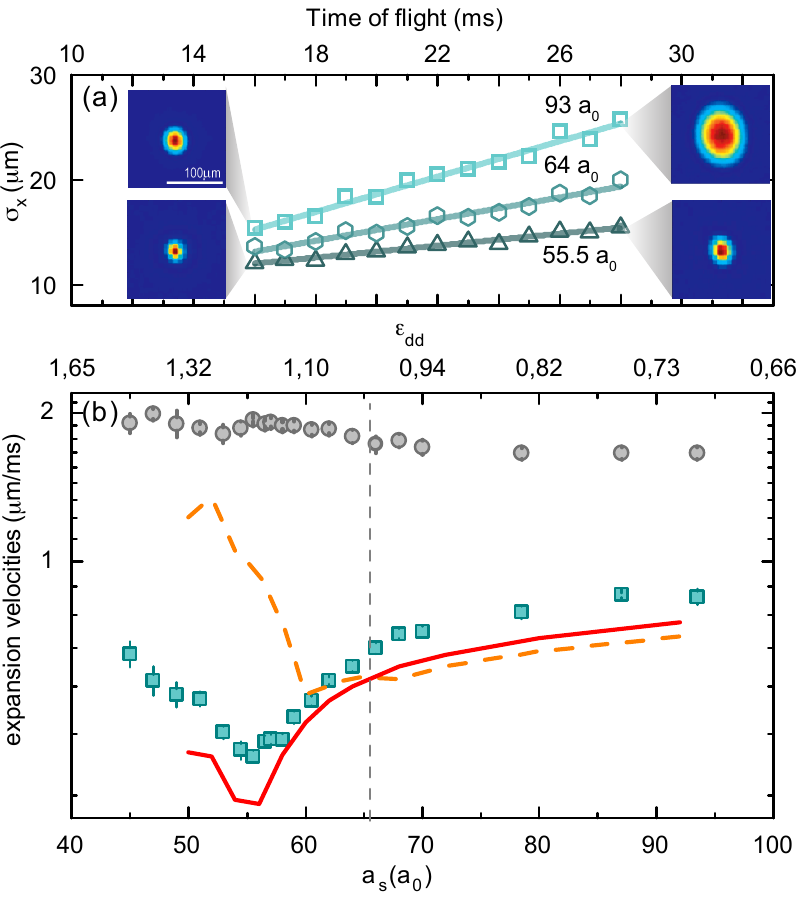}
\caption{\label{fig:expansion} (color online) {\bf Expansion dynamics across the BEC-to-droplet crossover}. 
(a) ToF evolution of the width, $\sigma_x$, of the high-density component for $\as= 93\,a_0$ (squares),  $64\,a_0$ (circles),  $55.5\,a_0$ (triangles). The lines are fit to the data using $\sigma_x(\ttof)=\sqrt{\sigma_{x,0}^2+v_x^2\ttof^2}$, from which we extract $v_x$. 
(b) $v_x$ as a function of $\as$ (squares). 
For comparison, the $\as$-independent expansion velocities of the thermal component are also shown (circles). The experimental data are in in very good agreement with our parameter-free theory from RTE simulations including LHY term (solid line) and rule out the MF scenario (dotted line). For clarity, we only show $v_x$, similar results are found for $v_y$. 
}
\end{figure}

 %
 Besides their unlike stability diagram, collective excitations, and density distribution, a dilute BEC in the MF regime and a quantum droplet are also expected to exhibit a markedly different expansion dynamics. While the first is confined by an external trapping potential and thus freely expands in its absence, a droplet state is self-bound by its underlying interaction in analogy with the He-droplet case \cite{Waechtler:2016, Bisset:2016, Waechtler:2016b, Baillie:2016}. As in our previous discussions, the evolution from a trap-bound to a self-bound solution is expected to be regulated by the interplay between QFs and 3B loss processes. 
 
 We investigate the expansion dynamics of our system for various $\as$. To preserve the high density of the coherent component, our measurements focus on short timescales with $\tra=10\,$ms and $\tho=5\,$ms. After preparing the system at the desired $\as$, we abruptly switch off the ODT, let the gas expand for a variable $\ttof$, and probe the cloud width using the $\parallelsum$-imaging. 
 We fit the observed density distribution to a double-Gaussian function, as previously described. To extract the width $\sigma_\eta$ of the high-density core ($n_{\rm core}$),  we compute the second moments $\sigma_\eta^2= \int \eta^2 n_{\rm hG}(\br)d\br$ along $\eta=x,z$. 
Figure \ref{fig:expansion}\,(a) exemplifies the ToF evolution of $\sigma_{\eta=x}$  at $\as=93 a_0$, $64 a_0$, and $55.5 a_0$. When entering the $\epsdd>1$ regime, atoms in the high-density core exhibits a  marked slowing down of the expansion dynamics, which can not be explained within the MF approach. 

To systematically study this effect, we repeat the above measurements for different values of $\as$ (i.\,e.\,$\epsdd$).
From $\sigma_x(\ttof)$,  we extract the value of the expansion velocity, $v_x$, by fitting the data to $\sigma_x(\ttof)=\sqrt{\sigma_{x,0}^2+v_x^2\ttof^2}$. 
Figure \ref{fig:expansion}\,(b) shows $v_{x}$  in an $\epsdd$ range from $0.7$ to $1.5$. When the system approach the droplet regime with decreasing scattering length ($\as < \add$), $v_\eta$ undergoes a strong reduction and drops to a minimum equal to $v_{x}=0.40(2)\,\um/{\rm ms}$ at  about $56\,a_0$ ($\epsdd\sim 1.17$). For further lowering of $\as$, $v_x$ starts to increase again. A similar behavior is observed for $v_y$.
We note that only the high-density component reveals this intriguing dependency on $\as$, whereas the thermal part shows a mainly constant expansion velocity.

Our observations agree well with the theory including the LHY term; see solid line in Fig.\,\ref{fig:expansion}(b). 
The ToF evolution  is calculated  using a multi-grid numerical scheme\,\cite{suppmat}. We record the evolution of $\sigma_\eta$ with $\ttof$ and extract the corresponding expansion velocities from the asymptotic behavior of ${{\rm d }\sigma_\eta}/{{\rm d} \ttof}$.
Our simulations show a slowing down with a minimum of $v_x=0.32\,\um/{\rm ms}$ at $\af\sim 56\,a_0$ ($\epsdd\sim 1.17$), followed by an increase at lower $\af$ \footnote{We note that the small offset between the theoretical and experimental data can be due to systematics effects in our imaging, e.\,g.\,final resolution of the imaging system and stray fields during ToF \cite{suppmat}.}. 
In contrast, calculations in absence of beyond-MF corrections fail to reproduce the experimental data. Here, the velocity is first more drastically reduced above the MF instability threshold $\epsdd \sim 1$ than it is expected with LHY corrections, it then already increases at this threshold. The first point relies on the trivial slowing down of a BEC whose mean repulsion energy is weakened (by reducing $\as$ or decreasing its population $\Nc$). The second point reveals a collapsing behavior that gives rise to an explosive evolution of the density profile. The minimal velocity is here found to be $v_x=0.56\,\um/{\rm ms}$ at $\af=68\,a_0$, which is a much higher value than both our experimental results and our theory predictions including the LHY correction.


The expansion behavior can be qualitatively well understood considering the so-called released, or internal energy, $\ER$.
This is the energy of the system when substracting the energy related with the confinement \cite{Pitaevskii:2003}.
In the MF scenario, $\ER>0$ as long as the ground state is stable. The BEC expands ballistically and $v_\eta$ decreases for decreasing $\as$ and $N$.
In the unstable regime, the expansion velocity depends crucially on the value of $\tho$ at which the trap is switched off due to the occurrence of an in-situ collapse dynamics.
Contrary, in the presence of QF, a stable ground state always exists. The sharp variability in $\tho$ is expected to be suppressed. Assuming a fixed $\Nc$~(\emph{i.e.} no 3B losses), one can show that 
$\ER$ decreases with decreasing $\as$ and eventually reaches $\ER<0$ for $\as<\aSB$, marking the onset of the self-bound (SB) solution~(e.g. $\aSB=56a_0$ for $N=1.2\times 10^5$) \cite{suppmat}. 
However, in stark contrast to the MF case,  $\ER$ increases with decreasing $\Nc$ in the droplet regime.  We note that $\aSB$ is then shifted to lower values when $\Nc$ gets reduced by 3B losses, thus affecting the lifetime of the self-bound solution.

The existence of a minimal expansion velocity is thus a direct consequence of the competition between the decrease of $\ER$ for decreasing $\as$ at a fixed $\Nc$, and the increase of $\ER$ for decreasing $\Nc$ in the droplet regime. In the crossover regime, the system smoothly evolves towards a fully self-bound state ($v_\eta=0$) until 3B losses, occurring in trap or in the initial phase of the expansion, set in to unbind the system and to reduce the lifetime of the droplet state.

\section{\label{Sec:Conclusion}Conclusion}

In summary, we have demonstrated the existence of the crossover from a dilute BEC to a quantum droplet state driven by QFs. Our experiments not only 
demonstrate that LHY stabilization is a general feature of strongly dipolar gases, but show as well an excellent quantitative agreement 
with our parameter-free theory, which is based on a generalized GP equation with LHY correction. The agreement concerns the  lowest-lying excitation mode, 
the evolution of the atom losses, and the ToF expansion. Concerning the latter, we observe a prominent reduction of the expansion velocity to a minimum value, which provides a fingerprint of the competition between 3B losses and LHY stabilization.


\section*{\label{Sec:Acknowledgement}Acknowledgement}

We thank  M.\,Baranov, R.\,van Bijnen, R.\,N.\,Bisset and B.\,Blakie, E.\,Demler, R.\,Grimm, T.\,Pfau and the Stuttgart Dy team for multiple useful discussions. We thank G.\,Faraoni for her support in the final stage of the experiment.   
The Innsbruck group thanks the European Research Council within the ERC Consolidator Grant RARE no. 681432. LC is supported within the Marie Curie Project DIPPHASE no. 706809 of the European Commission.
 The Hannover group thanks the support of the DFG (RTG 1729). Both groups acknowledge the support of the DFG/FWF (FOR 2247).

\appendix*
\section{Supplementary material}
\subsection{Bose-Einstein condensation of \Er}
We prepare an ultracold gas of the \Er\  isotope following a similar trapping and cooling scheme as the one employed for $^{168}$Er \cite{Aikawa:2012,Frisch:2012}. We load a crossed-ODT from a narrow-line MOT of $3\times10^7$ \Er\ atoms at about $10\mu{\rm K}$.
At the end of the MOT sequence, the atoms are automatically spin-polarized in their lowest Zeeman sub-level\,\cite{Frisch:2012}.
The dipole orientation follows the one of the external applied magnetic field, $\bB$. In our experiment, the latter is controlled by independent tuning of the components $B_x,B_y, B_z$ along the experimental coordinate system $(x,y,z)$, as defined in Fig.\,\ref{fig:setup} (lower inset). 

The ODT results from the crossing at their foci of two red-detuned laser beams at a wavelenght of  $1064\,$nm. One beam propagates horizontally along the $y$-axis, and the other propagates vertically and nearly collinear to the $z$-axis. 
The $z$-beam has a maximum power of 10\,W and an elliptical profile defined by its waists of $(w^{(z)}_x,w^{(z)}_y)=(110,55)\,\mu{\rm m}$. The $y$-beam has a maximum power of 27\,W, a vertical waist $w^{(y)}_z=18\mu{m}$, and a tunable horizontal waist, $w^{(y)}_x$. The latter can be conveniently tuned from $1.57\,w^{(y)}_z$ to $15\,w^{(y)}_z$ by time averaging the frequency of the first-order deflection of an Acousto-Optic Modulator (AOM). This scanning scheme enable both an efficient loading of the MOT into the ODT ($>30\%$ of the atoms are loaded) and an adiabatic and controlled tuning of the trap aspect ratio $\lambda$ over a broad range.
We achieve Bose-Einstein condensation of \Er\ atoms by means of  evaporative cooling  in the crossed ODT at 
$|\bB|=B_z =1.9\,$G ($\as =80(2)\,a_0$). 
Typically, we first rapidly (in 600\,ms) reduce the power and aspect ratio $w^{(y)}_x/w^{(y)}_z$  of the $y$-beam from 24\,W to 4\,W and $10$ to $1.6$, respectively. We further decrease the power of the $y$-beam from 4\,W to 0.3\,W in $3\,$s in an exponential manner and then exponentially increase the aspect ratio $w^{(y)}_x/w^{(y)}_z$ from 1.6 to 8 in 2.5\,s. The final trap frequencies are typically of $(\nu_x; \nu_y; \nu_z) \sim (40; 40; 180)\,$Hz. We finally obtain BECs of typically $N=1.2\times 10^5$ atoms with more than 80\% condensed fraction and a temperature $T\sim 70\,$nK. We typically measure $N$ and the condensed fraction from a bimodal fit of the 2D column density distribution measured along $\parallelsum$-imaging with $\ttof=27\,$ms. $T$ is extracted from the evolution of the thermal size of the bimodal fit with $\ttof$ varying from $14$ to $28$\,ms.

\subsection{Experimental setup and axes}
In our setup we define the orthonormal $(x,y,z)$-coordinate system in the following way: the vertical axis $z$ is aligned with gravity and the $y$ axis with the horizontal ODT beam; see Fig.\,\ref{fig:setup}. The polarizing magnetic field is created by three orthogonal pairs of coils. 
These pairs of coils define an orthonormal  $(X,Y,Z)$-coordinate system with $Z=z$ and $(X,Y)$ rotated by a small angle $\theta$ as compared to $(x,y)$. The magnetic field components $B_X,\ B_Y,\ B_Z$, each created by each pair of coils, can be controlled independently. We estimate $\theta$ to be about 15$\dg$ by sensing directly the atomic cloud, as its dipolar character makes it sensitive both to the trap geometry and to the magnetic field direction.

The small tilt $\theta$ between the dipoles and $y$ causes a small reduction of the mean DDI energy and corresponding small shifts of the expected characteristics compared to the one predicted for $\theta=0$: the MF collapse threshold should appear at a lower $\as$ and, for a given $\as$, $\nuax/\nu_{\parallelsum}$ and $v_\eta$ are shifted respectively down and up.
We have experimentally evaluated  the shift of $\nuax$ deep in the stable BEC regime ($|\bB | = \sqrt{B_X^2+B_Y^2}=2\,$G) to be of the order of 2\% and in the droplet regime to be about 10 to 15\%, as confirmed also by our theoretical predictions.

Finally, we also note a tilt between the $\parallelsum$-imaging beam and our reference frame, corresponding to an angle of $\theta^{\rm im}_{\parallelsum,0}\sim 28\dg$ compared to $y$ and $\theta^{\rm im}_{\parallelsum}=13\dg$ compared to $Y$ in the $xy$-plane.
The $\perp$-imaging axis is tilted by $\sim 15\dg$, mainly in the $xz$-plane.
Such tilts shift the observed size compared to the ideal case of imaging along and perpendicular to the dipoles.
Such an effect is not expected to impact the measurement of the collective frequencies, whereas it might bring a systematic shift of $v_x$ because of a mixed projection of $v_X$ and $v_Y$, the two first velocity components in the $(X,Y,Z)$-coordinate system, which are respectively perpendicular and along the dipoles.

In the theoretical calculations presented in the main text (Eq. \eqref{eq:H}, RTE and Gaussian ansatz), for simplicity, we do not account for these angles, whose effects are estimated to be smaller than our systematics.



\subsection{Precision measurements of the $\as$-to-$B$ conversion in a three-dimensional optical lattice}


Precise determination of the $\as$-to-$B$ conversion is a delicate issue, especially in the case of complex species, such as Er, for which comprehensive multi-channel calculations are still out of reach and the knowledge of $\as$ should thus rely on experimental investigations.
We perform lattice modulation spectroscopy in a three-dimensional optical lattice. From the measurements of the energy gap in the Mott insulator state we extract $\as(B)$.
Our lattice experiments focuses in the region of low $B$-field [$0,2.5$\,G] and are based on a lattice setup and procedure similar to the one described in Ref.\,\cite{Baier:2015}.
In brief, after producing the BEC, we load the atoms in a three-dimensional optical lattice by exponentially increasing the lattice-beam power in 150\,ms.
The typical  final depths are $(s_x,s_y,s_z)=(20,20,100)$, given in unit of the respective recoil energies $h\times (4.2,4.2,1.05)\,$kHz. At these lattice dephts, the gas is in the Mott insulator state. We then vary $B=(0,0,B_z)$ to the desired value by rapidly changing $B_z$ in 2\,ms, either just before or just after loading the lattice. We use the latter option for the smallest $B$ values at which $L_3$ is enhanced because of its proximity to the near-zero-field resonance.
In this case, we further hold 20\,ms to make sure the magnetic field is fully established before performing the modulation. 

To perform spectroscopy measurements, we sinusoidally modulate $s_y$ at a variable frequency $\nu_{ m}$ for 90\,ms with a peak-to-peak amplitude of about $30\%$. Finally, we ramp down the lattice depths to zero in 150\,ms, and measure the recovered condensed fraction as a function of $\nu_{\rm m}$ from $\parallelsum$-imaging ToF picture. For the smallest $B$ values considered, we also ramp $B$ back to 2\,G in 2\,ms before switching off the lattice-beams in order to again minimize 3B loss effects.

When varying $\nu_{\rm m}$, we observe a resonant depletion of the condensate due to particle-hole excitations. The resonance condition in the Mott-insulator regime is given by\,
\begin{equation}
h\nu_{\rm ex} = U_{\rm s} + U_{\rm dd}- V_{\rm dd}.
\end{equation}
Here $U_{\rm s}$, $U_{\rm dd}$ and $V_{\rm dd}$ are respectively the on-site contact interaction, the on-site dipolar interaction and the nearest-neighbor dipolar interaction along $y$ in the corresponding extended Bose-Hubbard model.  $U_{\rm dd}$ and $V_{\rm dd}$ can be accurately predicted from the knowledge of the lattice depths and dipole orientation and in our typical experimental condition, they are equal to $h \times -344.8\,$Hz and $h \times 31.5\,$Hz respectively. 
 By subtracting the theoretical dipolar contributions to the measured frequency, we extract $U_{\rm s}$, which is directly proportional to the scattering length $\as$.
A precise mapping of $\as$ in the ultralow $B$-field region is then obtained by repeating the above measurements at various $B$ values; see Fig.\,\ref{fig:setup}. 


In the region [$0,2.5$]\,G, our lattice spectroscopy reveals the presence of two FRs, one located at about zero $B$ field and the other one at about 3\,G. The existence and position of these two FRs agree with our Feshbach spectroscopy measurements performed in an harmonically trapped thermal cloud, where the maxima in 3B losses approximately pinpoint the resonance positions. In this measurement, further FRs are identified at $4.1\,$G and $5\,$G.

The scattering length of $^{166}$Er can be parametrized by the following simple expression \cite{Chin:2010}
\begin{equation}
\label{eq:a_to_b}
\as(B)=a_{\rm bg}(B)\left[1+ \sum_{i=1}^4\frac{\Delta B_i}{B-B_i}\right]
\end{equation}
in which the specific positions ($B_i$) and widths ($\Delta B_i$) of the two first resonances as well as the background scattering length are obtained from a fit to our lattice spectroscopy measurement. From the fit, we obtain $B_1=48(45)$\,mG, $\Delta B_1=39(20)\,$mG, $B_2=3.0(1)\,$G, $\Delta B_2=110(35)\,$mG. The $B$-dependent background scattering length $a_{bg}(B)$ accounts for the overlapping resonances and reads as  $a_{\rm bg}(B) =62(4)+k B$ with $k=5.8(1.2)\,a_0/{\rm G}$. We also account for the small effect of the two next resonances, whose positions $B_3,\ B_4$ and widths $\Delta B_3, \Delta B_4$ are fixed to their estimates from the loss-spectroscopy measurements to 10\,mG. We check that the precise values of this parameters has little effect on our empirical description along Eq.\,\ref{eq:a_to_b}.

 \subsection{Ramps in scattering length}
 \begin{figure}[b]
\includegraphics{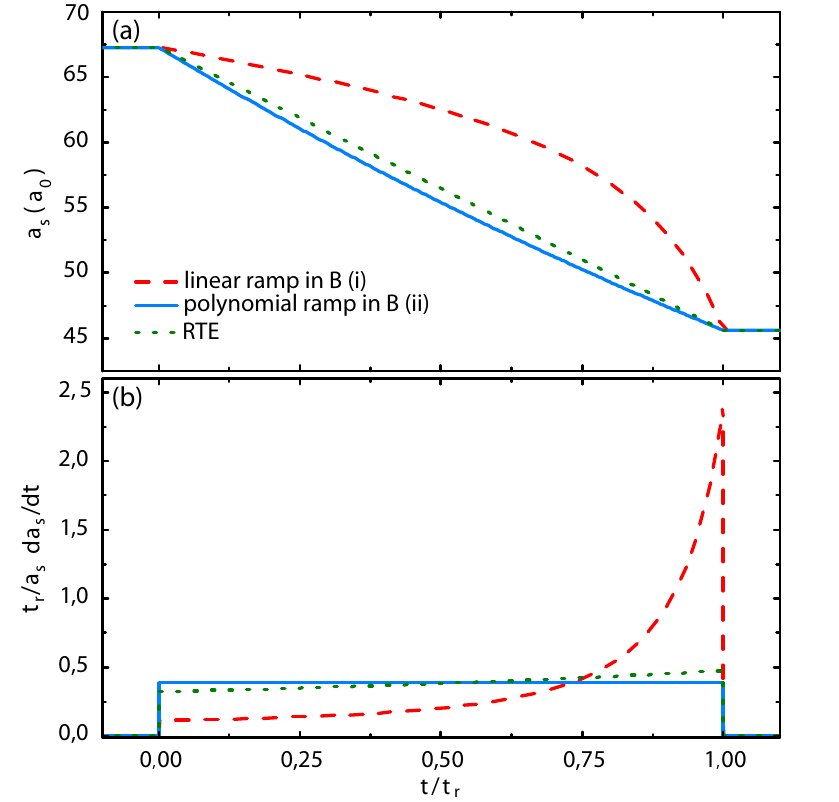}
\caption{\label{fig:S3_ramp} Predicted evolution of the scattering length $\as$ (upper panel) and the adiabaticity parameter
$\frac{1}{\as}\frac{{\rm d}\as}{{\rm d}t}/\nu_{\parallelsum}$
(lower panel) over the ramp from $B=0.8\,$G to the extreme $\Bf=0.17\,$G ($\af=45\,a_0$). The parameters are shown as a function of the normalized time $t/\tra$ and the universal variation of the adiabaticity parameter is obtained from normalizing to $\nu_{\parallelsum}\,\tra$. We show the cases of three different ramps: a linear ramp in $B$ (dashed red line) which is used in a first set of experiments (i), a polynomial ramp in $B$ (solid blue line) which is used in a second set of experiments (ii), a linear ramp in $\as$ (dotted green line) which is used in the simulation (RTE).}
\end{figure}
 Our measurements rely on controlled variations of the scattering length $\as(B)$.
 In our experiments, we either adiabatically change $\as$ using $\tra=45\,$ms or we quench it using $\tra=10\,$ms. The adiabatic condition for $\as$ reads as
 \begin{equation}
 \label{eq:adiabaticity}
 \frac{1}{\as}\frac{{\rm d}\as}{{\rm d}t}\leq \min\left(\nu_x,\nu_y,\nu_z\right)=\nu_{\parallelsum} \text{ for } \lambda\ll 1 
 \end{equation}
As shown in Fig.\,\ref{fig:S3_ramp}, we use two different types of time variations of $B$ and thus of $\as$: (i) a simple linear ramp in $B$ and (ii) we design a specific $B(t)$ variation in order to minimize the adiabaticity parameter
$\frac{1}{\as}\frac{{\rm d}\as}{{\rm d}t}/\nu_{\parallelsum}$.
The resulting $\as$ shows an exponential-type variation with $t$.
The adiabaticity condition of Eq.\,\eqref{eq:adiabaticity} is more stringent for lower $\as$. For ramping down to $\as=48\,a_0$, we find that (i) verifies Eq.\,\eqref{eq:adiabaticity}  for $\tra\gtrsim 100\,$ms and (ii) for $\tra\gtrsim 20\,$ms. 
Data from Figs.\,\ref{fig:setup}, \ref{fig:modes} and \ref{fig:losses}\,(b-d) (resp. Figs.\,\ref{fig:density_profile}, \ref{fig:losses}\,(a) and \ref{fig:expansion}) use ramp (i) (resp. (ii)).
 
 For our theoretical description, we use a linear change of $\as(t)$, similar to case (ii).

\subsection{Determination of the three-body recombination rate coefficient}
\begin{figure}[b]
\includegraphics{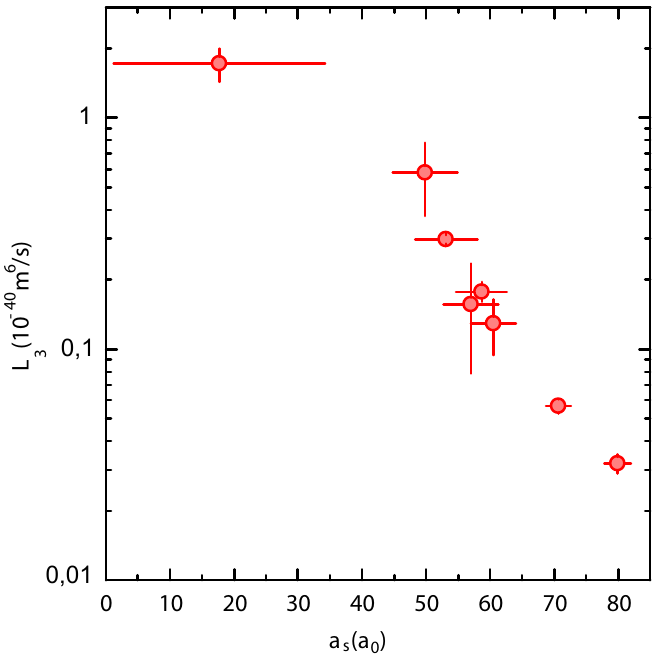}
\caption{\label{fig:S1_L3} Measured 3B recombination rate coefficient $L_3$ of a quantum degenerate gas of \Er\ as a function of $\as$ for $B$ varying from 0.1 to 1.9\,G. We measure $L_3^{\rm th}$ on a thermal gas at $T=490(10)\,$nK using Boltzmann law and taking into account anti-evaporation effect; see text. }
\end{figure}
Since three-body inelastic losses play a crucial role in the many-body dynamics and lifetime of the droplet state, we run a dedicated set of measurements to determine $L_3$.
We first prepare a non-degenerate thermal sample of Er atoms at $T=490(10)\,$nK in an harmonic trap. We then record the decay of the atom number, $\Nt$, as a function of $\tho$ in a range from 0 to 1\,s. $\Nt$ is obtained from a Gaussian fit to the measured density distribution. 

To fit the time evolution of $\Nt$, we use the integrated 3B rate equation, which reads as  $\frac{1}{\Nt}\frac{{\rm d}\Nt}{d\tho}=- L_3^{\rm th} \langle n^2 \rangle$. Here, $\langle n^2 \rangle$ is the mean square density on the cloud. To describe the scaling of $\langle n^2 \rangle$ with $\Nt$, we use its prediction for an ideal gas at thermal equilibrium at $T$ whose state occupancies follow Boltzmann law and take into account the anti-evaporation effect\,\cite{Weber:2003}. Then  $L_3^{\rm th}$ is extracted from a fit of $\Nt(\tho)$ along:
 \begin{equation}
 \Nt(\tho)=\frac{N_0}{(1+3\gamma_0N_0^2\tho)^{1/3}}
 \end{equation}
 where $N_0$ is the atom number at $\tho=0\,$ms and $\gamma_0$ is defined via the relation
 \begin{equation}
 L_3^{\rm th}= \sqrt{27}\,\gamma_0\left(\frac{k_{\rm B} T}{2\pi m \bar{\nu}}\right)
 \end{equation}
 with $k_{\rm B}$ the Boltzmann constant and $\bar{\nu}=(\nu_x \nu_y \nu_z)^{1/3}$.
 
 We account for the $\as$-dependence of $L_3^{\rm th}$ near a FR by repeating the measurement at different $B$. We check that the measured $L_3^{\rm th}$ does not depend on the $\bB$ orientation and measure its B-dependency using $|\bB|=B_y$ and for $B$ varying in 0.1 to 1.9\,G. 
Note that, due to the bosonization effect, the $L_3$ in a quantum degenerate bosonic gas is equal to $L_3^{\rm th}/3$. 
Figure \ref{fig:S1_L3} shows  $L_3$ in a quantum degenerate bosonic gas of $\Er$ as a function of $\as$ using the measured $\as$-to-$B$ conversion. Despite the existence of many coupled molecular potential in Er, we measure a comparable low $L_3(\as)$ as compared to the typical values reported in alkali atoms. $L_3(\as)$ varies between $1.7(3)\times 10^{-40}$m$^6/$s at $\as=18(17)$ and $3.2(3)\times 10^{-42}\,$m$^6/$s at $\as=80(2)\,a_0$.

\subsection{Time-of-Flight measurements}
For our ToF measurements, we abruptly extinguish the ODT in about $100\,\mu{\rm s}$.
In order to accurately image our gas while minimally influencing its dynamics during the expansion, we split the ToF in two parts. During a first part, lasting $\ttof-t_{\rm B}$, $\bB$ remains unchanged and the dynamics occur at the original $\as$ and dipole orientation. At time $t_{\rm B}$ before the image is taken, $\bB$ is modified both in amplitude and in direction, in order to correctly set the quantization axis for the imaging light to be $\sigma_-$ polarized. The amplitude $|\bB|$ of the imaging field is chosen constant and equal to 0.31\,G for all  $\as$ considered. $t_{\rm B}$ is set to be 12\,ms for $\parallelsum$-imaging and 15\,ms for $\perp$-imaging where the change in $\bB$ is more drastic for the typical dipole orientation ($Y$) used in this experiment.

We note that our resolution limit for both $\parallelsum$- and $\perp$-imaging is estimated to be $\gtrsim 3\,\mu{\rm m}$. Moreover the effective pixel sizes are set to $8.4$\,$\mu{\rm m}$ and $3.1$\,$\mu{\rm m}$ in our setup. These limit the size of the structure we are able to observe as well as the minimal $\ttof$ we can use, which is typically $\ttof\geq16\,$ms.



\subsection{Averaging experimental density distribution}
 In order to obtain a better image quality and resolution, we typically average four experimental absorption images taken in the same condition and with the same experimental series (\emph{i.e.} within less than a few hours interval).
 In order to average the images we first define a region of interest (ROI) of typically $300\,\mu{\rm m}\times300\,\mu{\rm m}$ containing the cloud shadow and translate the ROI to superimpose the cloud centers. To estimate the translation amplitudes for each individual image, we use the center from a simple Gaussian fit to the 2D density distribution ROIs.
 In this averaging process, we use a sub-grid resolution of $1/10$ of a pixel to more accurately superimpose the centers. We fit the averaged density distribution after binning back to the original resolution. We note that fits on the sub-pixel resolved images give similar results.

\subsection{Extracting the frequency of the collective modes}

As stated in the main text, we focus on the axial mode, which reveals itself by a  collective oscillation of the axial size $R_{\parallelsum}$ of the BEC, along with smaller amplitude oscillations of the radial size in phase opposition. We extract its frequency $\nuax$ by studying the larger amplitude oscillation of $R_{\parallelsum}$. For this, we probe the ToF density distribution of the gas with $\perp$-imaging after a ToF of 30$\,$ms. We focus on the sizes of the central, high-density component of the cloud, which we study as a function of $\tho$ for different $\as$. We note that the precise shape of the column density profile is expected to change as a function of $\as$ and this in a different way for the two axis $x$ and $y$ under observation in $\perp$-imaging. This complicates the analysis, in particular compared to the $\parallelsum$-imaging where both axis are nearly equivalent. Here, we extract the sizes of the central component, using various methods and select the most reliable method according to $\as$. Typically, we use a bimodal MF TF plus Gauss fit for $\as\geq 57\,a_0$. For $\as\leq 57\,a_0$ we select a central region in the cloud and perform a simple Gaussian fit on it.  Such a determination should give access to the variations, if not to its absolute value, of $R_{\parallelsum}$ with $\tho$ at fixed $\as$  and thus enable to determine $\nuax$.  

To fit $\nuax$, we use a damped-sine function of $\tho$. Typically we fit the evolution of $R_{\parallelsum}$ for $\tho$ varying from 0 to few hundreds of ms, depending on the damping rate observed. The upper value of $\tho$ used is never less than 150\,ms such that at least 4 to 5 oscillations are observed and fitted. We also note that for our shortest $\tra=10\,$ms we typically discard the first 4\,ms of the evolution in order to ensure that the magnetic field is safely stabilized at its target value.



\subsection{Bimodal fits of the density distribution in $\parallelsum$-imaging.}
To quantitatively analyse the experimental column density distribution imaged along $\parallelsum$-imaging, we perform bimodal fits on the 2D averaged profiles $n_{\parallelsum}(x,z)$. The bimodal fits are made of the sum of two peak distributions, describing respectively a high-density, coherent part and a thermal incoherent background. To account for the change of the profile of the density distribution across the BEC-to-droplet crossover, we use two types of fitting functions $f_{\rm fit}(x,z)$:
(i) 
A sum of a MF TF and a Gaussian distribution, which  account respectively for the coherent and the thermal part. For the integrated column density, the MF TF distribution writes $f_{\rm TF}(x,z) =  \left(1-\frac{(x-x_{0})^2}{R_x^2}-\frac{(z-z_{0})^2}{R_z^2}\right)^{3/2}$\,\cite{Pitaevskii:2003}.%
(ii) 
The sum of two Gaussian distributions. Typically the central Gaussian can be anisotropic with any orientation axis.

As expected, the thermal background is broader than the central coherent component and for the $\ttof$ considered its width is typically at least $1.6$ times larger. It is first fitted by taking out the central part of the density distribution at radius typically smaller than 1.3 times its width. 

The quality of the bimodal fit is evaluated by the norm of the residue $\rho=1-\frac{\int {\rm d}x\,{\rm d}z\,\left(n_{\parallelsum}(x,z)-f_{\rm fit}(x,z)\right)^2}{\int {\rm d}x\,{\rm d}z \,n_{\parallelsum}(x,z)^2}$. For both distributions, it satisfies $\rho>0.98$.

\subsection{Describing the atom number decay}

\begin{figure}[b]
\includegraphics{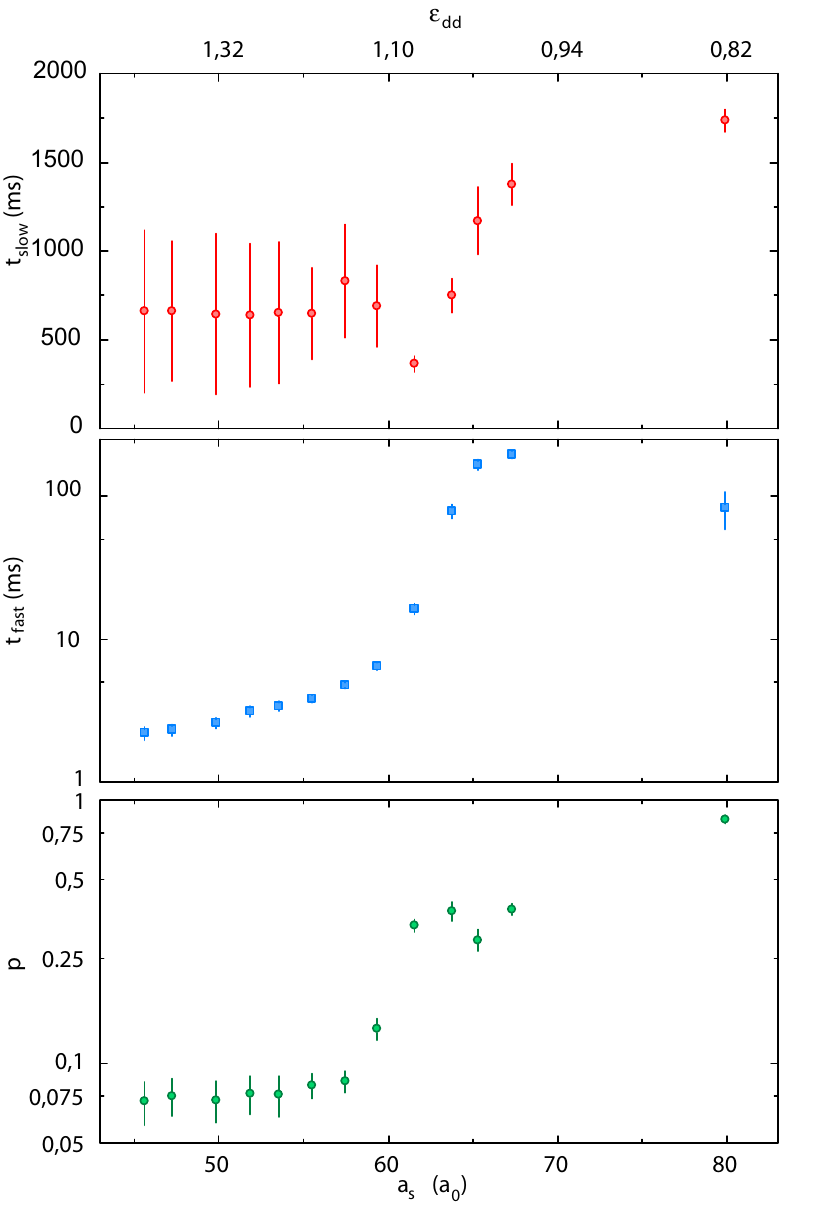}
\caption{\label{fig:S2_dec} Fit parameters for the evolution of  $\Nc(\tho)$. 
}
\end{figure}

In Figure \ref{fig:losses}\,(c) and in the main text we have briefly described the evolution of the atom number in the coherent part $\Nc$ with $\tho$. There our aim was to extract the mean density and we did not expand on the mere description of the $\Nc(\tho)$. 

We note that $\Nc(\tho)$ is well accounted for by a double exponential decay evolution of respective amplitude $N_0(1-p)$ and $N_0p$. $N_0$ is the initial atom number. Each decay corresponds to a different time constant, respectively  $\tfa$ and $\tsl$, and starts after a different delay time, respectively $\td$ and $\tD$. We fix $\tD=\td+2\tsl$. For $\tra=10\,$ms $\td$ is approximately constant and equal to $3.65\,$ms. The absence of evolution for $\tho<\td$ indicates that the magnetic field may not have reached the target value in the first ms. For $\tra=45\,$ms, $\td$ can be set to 0. $\tfa$, $\tsl$ and $p$ (and $N_0$) depend on $\as$, typically decreasing with it. The evolution is illustrated in Fig.\,\ref{fig:S2_dec} for $\tra=10\,$ms.

\subsection{Gaussian Ansatz including $L_3$}
A good qualitative (and to a large extent quantitative) insight in the physics of dipolar condensates in the presence of LHY stabilization 
may be gained from a simplified Gaussian ansatz of the form
\begin{equation}
 \psi({\mathbf r},t)= \sqrt{N(t)}e^{\mathrm{i}\phi(t)}  \prod_{\eta=x,y,z} 
\frac{ {\rm e}^{-\frac{\eta^2}{2w_\eta^2}+i\eta^2 \beta_{\eta}(t)}}{\pi^{1/4} w_\eta^{1/2}}, 
\label{GaussianAnsatz}
\end{equation}
where the variational parameters are the number of atoms $N(t)$, the global phase $\phi(t)$, the widths $w_\eta$, and the phase curvatures $\beta_\eta$. 
The Lagrangian density reads:
\begin{align}
 \mathcal{L}=&\frac{i \hbar}{2}\left(\psi \frac{\partial \psi^{*}({\bf r},t)}{\partial t}-\psi^{*}\frac{\partial \psi({\bf r},t)}{\partial t}\right)+\frac{\hbar^2}{2m}|\nabla \psi({\bf r},t)|^2 \nonumber \\ 
 &+V({\bf r})|\psi({\bf r},t)|^2+\frac{g}{2}|\psi({\bf r},t)|^4+\frac{2}{5}g_{\rm LHY}|\psi({\bf r},t)|^5\nonumber \\
 &+\frac{1}{2}\int d^3r' V_{\rm dd}({\bf r}-{\bf r}')|\psi({\bf r},t)|^2|\psi({\bf r}',t)|^2. 
 \label{LagrangianDensity}
\end{align}
We insert ansatz~\eqref{GaussianAnsatz} into Eq.~\eqref{LagrangianDensity}, obtaining the Lagrangian $L=\int d^3r \mathcal{L}$:
\begin{eqnarray}
L&=&N\left \{ \hbar\dot\phi+\frac{\hbar}{2}\sum_\eta \dot\beta w_\eta^2 +\frac{m}{4}\sum_\eta \omega_\eta^2 w_\eta^2 \right\delimiter 0 \nonumber \\
&+&2\left\delimiter 0 \frac{\hbar^2}{2m}\sum_\eta\left ( \beta_\eta^2 w _\eta^2+\frac{1}{4w_\eta^2} \right ) \right \} \nonumber \\
&+& N^2 \left \{ \frac{g ( 1+\epsilon_{dd} F (w_y/w_x, w_y/w_z ) )}{2(2\pi)^{3/2}\prod_\eta w_\eta}  \right \} \nonumber \\
&+& N^{5/2} \left \{ \left (\frac{2}{5}\right )^{5/2}\frac{g_{LHY}}{\pi^{9/4}\prod_\eta w_\eta^{3(2}} \right \},
\end{eqnarray}
with
\begin{eqnarray}
&&F(\kappa_x,\kappa_z)=\frac{1}{4\pi}\int_0^\pi d\theta \sin\theta \int_0^{2\pi} d\phi \nonumber \\
&&\left [ \frac{3\cos^2\theta}{\left ( \kappa_x^2\cos^2\phi+\kappa_z^2\sin^2\phi \right) \sin^2\theta+\cos^2\theta}-1 \right ].
\end{eqnarray}
The variational parameters are determined from the corresponding Euler-Lagrange equations~\cite{Filho2001}:
\begin{equation}
\frac{d}{dt}\left ( \frac{\partial L}{\partial\dot\lambda} \right )-\frac{\partial L}{\partial\lambda}=\int d^3r \left [ \Gamma \frac{\partial\psi^*}{\partial\lambda}+ \Gamma^* \frac{\partial\psi}{\partial\lambda}\right ],
\end{equation}
where $\lambda=N,\phi,w_\eta,\beta_\eta$, and $\Gamma({\mathbf r})=-i\hbar\frac{L_3}{2}|\psi({\mathbf r})|^4\psi({\mathbf r})$.
Introducing the dimensionless units $\tau=\tilde{\omega}t $, $w_\eta=\tilde{l}v_\eta$, $\tilde{l}=\sqrt{\hbar/m \tilde{\omega}}$, with $\tilde\omega=(\prod\omega_\eta)^{1/3}$, 
and after some algebra, we obtain a close set of equations for the Gaussian widths and the number of atoms:
\begin{equation}
\dot N=-\frac{3R}{\prod_\eta v_\eta^2}N^3, \label{eq:Ndot}
\end{equation}
\begin{equation}
\ddot v_\eta=-v_\eta \left [ \frac{7R^2N^4}{\prod_{\eta'} v_{\eta'}^4}
+\frac{2RN^2}{\prod_{\eta'}v_{\eta'}^2}\sum_{\eta''\neq \eta}\frac{\dot v_{\eta''}}{v_{\eta''}}
\right ]-\frac{\partial U}{\partial v_\eta},\label{eq:vdot}
\end{equation}
%
with $R=\frac{L_3}{\pi^3 3^{5/2} \tilde\omega \tilde l^6}$, and 
\begin{eqnarray}
 U&=& \frac{1}{2}\sum_{\eta}\left [ v_\eta^{-2}+\left ( \frac{\omega_\eta}{\tilde\omega} \right )^2v_\eta^2 \right ]+\frac{2}{3}\frac{PQN^{3/2}}{\left(\prod_\eta v_\eta \right)^{\frac{3}{2}}}\nonumber \\
 &+&  \frac{PN}{\prod_\eta v_\eta} \left ( 1+\epsilon_{dd}F \left (\frac{v_y}{v_x},\frac{v_y}{v_z} \right )\right ),
 \label{Energy_vi}
\end{eqnarray}
with  $P=\sqrt{\frac{2}{\pi}}\frac{a}{\tilde{l}}$ and 
 $Q=\frac{512\, F(\epsilon_{dd})}{25\sqrt{5}\pi^{\frac{5}{4}}} ( a/\tilde l\,) ^{3/2}$.

Due to its simplicity, Eqs.~\eqref{eq:Ndot} and~\eqref{eq:vdot} permit a much more flexible simulation of the exact experimental conditions and sequences compared to 
the obviously more exact but numerically much more cumbersome simulation of the gNLNLSE. We have checked that the results of the Gaussian ansatz approach are in excellent 
agreement both qualitative and to a large extent also quantitative to full simulations of the gNLNLSE, in what concerns lowest-lying excitations, atom losses, and 
expansion velocities. 
%
%

\subsection{Self-bound droplets}
\begin{figure}[t]
\includegraphics[width=1\linewidth]{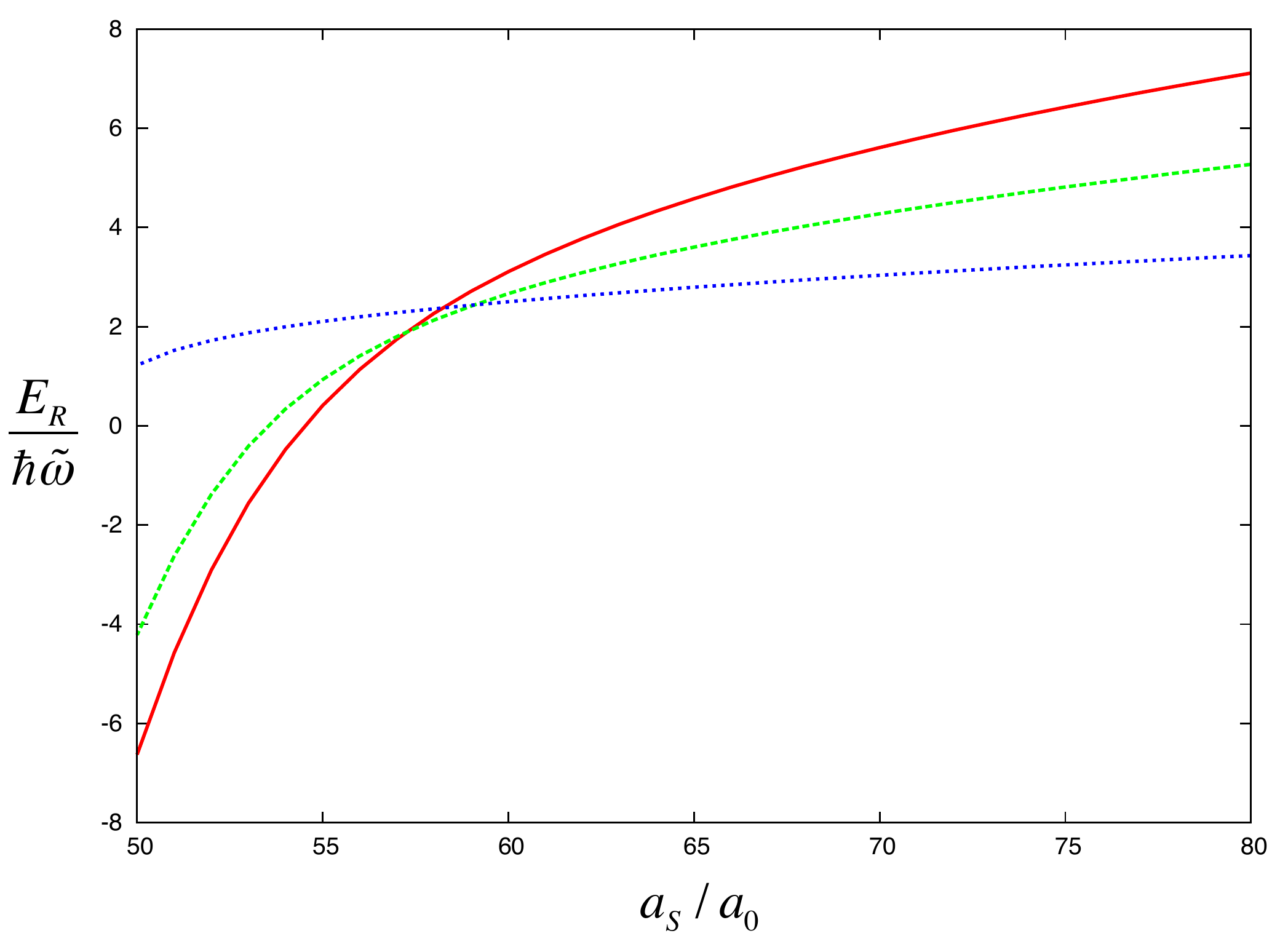}
\caption{Released energy $E_R$ as a function of $a_s$ for $N=1.2\times 10^5$~(solid), $5\times 10^4$~(dashed), and 
$10^4$~(dotted). Note that $E_R<0$ indicates a SB solution.}
\label{fig:S3} 
\end{figure}

The Gaussian ansatz approach allows for an intuitive understanding of the degree of self-bound~(SB) character of the system. As mentioned in the main text, 
a SB solution is characterized by a negative released energy, $\ER<0$. In absence of losses, we may evaluate $\ER$ by means of the Gaussian Ansatz  for the ground-state 
of a trapped BEC with scattering length $\as$ and  fixed $N$. Figure~\ref{fig:S3} shows the results for $\ER$ for different $N$  values. Whereas $\ER$ increases with growing $N$ for large $\as$, for small $\as$ in the droplet regime 
$\ER$ increases with decreasing $N$. For each $N$ there is a
finite scattering length $\aSB$ such that if $\as\leq \aSB$ the droplet will be fully self-bound ($v_\eta=0$). Given its $N$-dependence, $\aSB$ shifts to lower values with decreasing $N$. 
3B losses add a time dependence on $N$ and thus on $\aSB$ that governs the lifetime of the droplet state. We note that for small $(\tra,\tho)$ $\ER$ may change its sign during the ToF due to atom losses, i.e. a SB solution may unbind during the ToF.
In our experiments, the interplay of losses and LHY stabilization leads to a minimal released energy, that translates into a minimal expansion velocity as shown in the main text.

\subsection{Simulation of the ToF expansion using the gNLNLSE}
ToF expansion is simulated using the gNLNLSE by means of a multi-grid method, i.e. dynamically enlarging the numerical grid following the expansion. This is necessary due to the 
obvious difference in length scales at the beginning and at the end of the ToF. We note that the precise description of the ToF dynamics is very relevant, since in contrast to standard cases, nonlinear dynamics and losses here may play an important role during the expansion, especially within the LHY stabilized regime.



%

\end{document}